\documentclass{epl}
\usepackage{graphics,amssymb,amsmath}

\title{Dynamic model for failures in biological systems}

\author{J. Choi\inst{1} \and M.Y. Choi\inst{2,3} \and
B.-G. Yoon\inst{4}\thanks{Author to whom correspondence should be addressed.
e-mail: \email{bgyoon@ulsan.ac.kr}}}
\institute{
  \inst{1} Department of Physics, Keimyung University, Taegu 704-701, Korea\\
  \inst{2} Department of Physics, Seoul National University, Seoul 151-747, Korea\\
  \inst{3} Korea Institute for Advanced Study, Seoul 130-722, Korea\\
  \inst{4} Department of Physics, University of Ulsan, Ulsan 680-749, Korea
}

\begin{document}

\maketitle

\begin{abstract}
A dynamic model for failures in biological organisms is proposed
and studied both analytically and numerically.  Each cell in the
organism becomes dead under sufficiently strong stress, and is then 
allowed to be healed with some probability.  It is found that
unlike the case of no healing, the organism in general does not
completely break down even in the presence of noise.  Revealed is
the characteristic time evolution that the system tends to resist
the stress longer than the system without healing, followed
by sudden breakdown with some fraction of cells surviving.
When the noise is weak, the critical stress beyond which the
system breaks down increases rapidly as the healing parameter is
raised from zero, indicative of the importance of healing in
biological systems.
\end{abstract}

Many systems in nature under steady external driving are led into
failures. Examples are diverse ranging from disordered
heterogeneous materials, earthquakes, and even to social
processes.  The fiber bundle models, describing well these
systems, have received considerable attention~\cite{rev}. Most
studies of fiber bundles are based on the recursive breaking
dynamics at discrete time step. One of the typical features in the
models is that local stress arising from external driving tends to
produce avalanches of microscopic failures, resulting in
stationary macroscopic breakdown of the system. The main issue
here is thus the life time of a fiber bundle and the avalanche
size distribution under various conditions~\cite{hd,asl,lt}. There
are shortcomings, however, in discrete dynamics: It cannot
describe the time evolution in real continuous time, for
which a remedy has been suggested recently~\cite{ccy}.

Failures are also observed in biological systems, as a consequence
of disease for instance; these are different from other failures
in the sense that there may exist healing, in reparative or
regenerative ways. Here the evolution to the stationary state
appears more important than the stationary state itself.

In this Letter we extend our previous work on the dynamic model
for failures in fiber bundles~\cite{ccy} to biological systems
consisting of cells.  Incorporating the effects of healing into
the model in a manner similar to that used for neural networks~\cite{myc}, 
we derive equations of motion for biological organisms in the form of
delay-differential equations. From this formulation, we obtain the
evolution equation for the average number of living cells and find
that there generally remains a finite fraction of living cells in
the stationary state, manifesting the healing effects.  This
persists in the presence of noise, in sharp contrast to the case
of no healing.
The system is then explored numerically, which shows that
the presence of healing assists the system to resist the stress
longer before abrupt breakdown into the stationary state as well
as increases the critical stress of external load that brings
about the breakdown. The resulting behavior is reminiscent of the
time course of degenerative disease progression such as diabetes,
Alzheimer's disease, and possibly AIDS. We thus believe our
approach may serve as a prototype model for disease progression at
the cellular level, for which only a phenomenological description
is available so far~\cite{hp}.

We consider an organism consisting of $N$ cells under external
stress characterized by load $F$.  
Each cell has its own tolerance and endures the stress
below the tolerance, thus remaining alive. The cell may become
dead, however, if the tolerance is exceeded. We assign ``spin''
variables to these in such a way that $s_i=-1\, (+1)$ for the
$i$th cell alive (dead). The state of the organism is then
described by the configuration of all the cells, $\mathbf{s}
\equiv (s_1, s_2,..., s_N)$.  The total number $N_-$ of living
cells is related with the average spin $\bar s \equiv N^{-1}
\sum_j s_j$ via
\begin{equation}\label{intact}
N_{-}=\sum_{j=1}^N \frac{1-s_j}{2}=\frac{N}{2} (1- \bar s),
\end{equation}
and we are interested in how $N_-$ evolves in time as well as its
stationary value.  The total stress on the $i$th cell can then be
written in the form
\begin{equation} \label{f}
 \eta_i = f + \sum_j V_{ij}\,\frac{1+s_j }{2},
\end{equation}
where $f=F/N$ is the stress directly due to the external load and
$V_{ij}$ represents the stress transferred from the $j$th cell (in
case that it is dead). 
The death of the $i$th cell with tolerance $h_i$ is determined
according to:
\begin{eqnarray*}
& \eta_i < h_i ~\Rightarrow~ s_i=-1 \nonumber\\& \ \eta_i > h_i
~\Rightarrow~ s_i=+1 ,\label{dyn}
\end{eqnarray*}
which, in terms of the local field $E_i \equiv (\eta_i - h_i
)(1-\bar{s})/2 $, 
can be simplified as
  $ s_i E_i > 0$.
This determines the stationary configuration at which the system
eventually arrives.

For a more realistic description of the time evolution, we also
take into consideration the uncertainty (``noise'') present in
real situations, which may arise from imperfections, random
variations, and other environmental influences. We thus begin with
the conditional probability that the $i$th cell is dead at time
$t{+}\delta t$, given that it is alive at time $t$. For
sufficiently small $\delta t$, we may write~\cite{myc}
\begin{equation}\label{con1}
    p(s_i{=}+1, t{+}\delta t | s_i{=}-1, t; \mathbf{s}', t{-}t_d)
        =\frac{\delta t}{2t_r}\left[1+\tanh \beta E'_i \right],
\end{equation}
where $\mathbf{s}' \equiv (s'_1, s'_2, \ldots, s'_N )$ represents
the configuration of the system at time $t{-}t_d$ and $E'_i =
(\eta_i - h_i )(1-\bar{s'})/2 $ 
is the local field at time $t{-}t_d$. Note the two time scales
$t_d$ and $t_r$ here: $t_d$ denotes the time delay during which
the stress is redistributed among cells while the refractory
period $t_r$ sets the relaxation time (or life time). The
``temperature" $T\equiv \beta^{-1}$ measures the width of the
tolerance region of the cells or the noise level: In the noiseless
limit $(T =0)$ the factor $(1+\tanh \beta E'_i)/2$ in eq.
(\ref{con1}) reduces to the step function $\theta (E'_i)$,
yielding the stationary-state condition. 
We also assign the non-zero conditional
probability of the $i$th cell being repaired (regenerated) given
that it is dead at time $t$, according to
\begin{equation}\label{con2}
    p(s_i{=}-1, t{+}\delta t | s_i{=}+1, t; \mathbf{s}', t{-}t_d)
      = \frac{\delta t}{t_0},
\end{equation}
where $t_0$ is the time necessary for cell regeneration. Equations
(\ref{con1}) and (\ref{con2}) can be combined to give a general
expression for the conditional probability
\begin{equation} \label{cong}
p(-s_i, t{+}\delta t | s_i, t; \mathbf{s}', t{-}t_d)
  = w_i(s_i; \mathbf{s}', t{-}t_d )\delta t
\end{equation}
with the transition rate
\begin{equation}\label{tran}
w_i(s_i; \mathbf{s}', t{-}t_d )
      = \frac{1}{2t_r} \left[\left(a+\frac{1}{2}\right)+\left(a-\frac{1}{2}\right)s_i
        +\frac{1-s_i}{2}\tanh \beta E'_i  \right],
\end{equation}
where the healing parameter $a \equiv t_r /t_0$ measures the
relative time scale of relaxation and regeneration of a cell.
Here we point out that $t_d$, $t_r$, $t_0$, and accordingly $a$
may be complicated functions of system properties like the average
spin $\bar s$ and others, 
which may be incorporated into the model.  In this work we restrict
ourselves to the simplest case of these parameters being fixed.

The behavior of the organism is then governed by the master
equation, which describes the evolution of the joint probability
$P(\mathbf{s},t; \mathbf{s}', t{-}t_d)$ that the system is in
state $\mathbf{s}'$ at time $t{-}t_d$ and in state $\mathbf{s}$ at
time $t$.  Following the procedures in Ref.~\cite{ccy}, we rescale
time $t$ in units of the delay time $t_d$ and write accordingly the transition rate
 $w_i(s_i; \mathbf{s}')\equiv t_d w_i(s_i; \mathbf{s}', t{-}t_d)$,
which yields, in the limit $\delta t \rightarrow 0 $,
\begin{equation}\label{mast}
  \frac{d}{dt} P(\mathbf{s},t; \mathbf{s}', t{-}1)
 = -\sum_i [w_i(s_i; \mathbf{s}') P(\mathbf{s}, t; \mathbf{s}', t{-}1)
    - w_i(-s_i; \mathbf{s}') P(F_i\mathbf{s}, t; \mathbf{s}', t{-}1)]
\end{equation}
with $F_i\mathbf{s}\equiv (s_1, s_2,..., -s_i, s_{i+1},..., s_N)$.
Here it has been noted that contributions from the intermediate configurations
$\mathbf{s}' $ differing from $\mathbf{s}$ by only one cell survive;
those from other configurations are of order $(\delta t)^2$ or higher
and thus vanish in the limit $\delta t \rightarrow 0$.
Then equations describing the time evolution of relevant physical quantities,
with the average taken over $P(\mathbf{s},t; \mathbf{s}', t{-}1)$,
in general assume the form of differential-difference equations due to
the delay in the stress redistribution. In particular, the average spin for the $k$th cell,
$ m_k(t)\equiv \langle s_k \rangle \equiv\sum_{\mathbf{s},\mathbf{s}'}s_k
P(\mathbf{s},t;\mathbf{s}', t{-}1)$
satisfies
\begin{equation}
\tau \frac{d}{dt}m_k =
 \left(\frac{1}{2}-a\right)-\left(\frac{1}{2}+a\right)m_k+
    \left\langle\frac{\displaystyle {1-s_k}}{\displaystyle 2}\tanh \beta E'_k \right\rangle
  \label{eqn}
\end{equation}
where $\tau\equiv t_r/t_d$ gives the relaxation time (in units of $t_d$).

To proceed further, we assume equal load sharing, i.e., that the
stress is distributed to every cell uniformly.  In this case we
have $V_{ij}= \eta_j /N_-$ and accordingly, $(1-\bar s)\eta_i =
2f$ from eq. (\ref{f}). The infinite-range nature of equal
load-sharing allows one to replace $E'_k$ by its average $\langle
E'_k\rangle = f - (h_k /2)[1-\bar m(t{-}1)]$,
where it has been noted that $s'$ is the configuration at time
$t{-}1$, i.e., $\langle\bar s'\rangle =N^{-1}\sum_{j}\langle\bar
    s'_j \rangle =N^{-1}\sum_{j}m_j (t{-}1) \equiv \bar m(t{-}1)$.
For convenience, we now rewrite eq. (\ref{eqn}) in terms of the
average number of living cells at time $t$. Defining $x_k \equiv
(1-m_k)/2 $ and $\bar x\equiv N^{-1}\sum_{k}x_k = (1-\bar m)/2$,
we have, from eq. (\ref{intact}),
$ \langle N_{-}\rangle =
N \bar x $
and thus obtain from eq. (\ref{eqn}) the equation of motion for
the average fraction of living cells:
\begin{equation}\label{xk}
 \tau \frac{d}{dt}x_k (t)= a - \left(\frac{1}{2} +a\right) x_k (t)
             -\frac{1}{2} x_k(t)\tanh \beta [f {-}h_k \bar x(t{-}1)].
\end{equation}

We first examine the stationary solution of eq. (\ref{xk}) with
$dx_k /dt =0$:
%
\begin{equation} \label{}
    x_k =\frac{2a}{(1+2a)+ \tanh \beta (f{-}\bar x h_k )},
\end{equation}
which, upon averaging over the tolerance distribution $g(h)$, leads to the
self-consistency equation for the average fraction of living cells
or the ``health status'' of the organism
\begin{equation}\label{sta}
    \bar x =\int dh g(h)\frac{2a}{(1+2a)+ \tanh \beta (f- \bar x h )}.
\end{equation}
It is obvious that regardless of noise, the complete breakdown,
described by the null solution $\bar x =0$, is possible only for
$a=0$, i.e., when cells are not regenerated or repaired. This
contrasts with the case $a=0$, where noise usually gives rise to
the breakdown of the system ($\bar x=0$)~\cite{ccy,pc}. It is thus
concluded that the biological systems become robust against noise
due to the cell regeneration effects.

In the noiseless limit ($T=0$) we replace the term $\tanh
\beta(f{-}\bar x h)$ in eq. (\ref{sta}) by $\theta(f{-}\bar x h)-
\theta (\bar x h {-}f)$, to obtain
\begin{equation}\label{solbar}
    \bar x=\int dh g(h) \left[\theta (h \bar x {-}f)+\frac{a}{1+a}\;\theta(f{-}h \bar x)\right].
\end{equation}
%
For a continuous distribution of the tolerance, we formally solve
eq. (\ref{solbar}) by simply performing the integration
\begin{equation}\label{G}
    \bar x =1- \frac{1}{1+a}\;G(f/\bar x),
\end{equation}
where $G(h)=\int_0^h dh' g(h')$ is the cumulative distribution of the tolerance.
Equation (\ref{G}) shows that a majority of cells in general
becomes dead at stress above a critical value $f_c$, which depends
on $a$.
This can be manifested with simple distributions, e.g., the bimodal distribution
$g(h)=\rho \delta (h{-}f_1) +(1-\rho)\delta (h{-}f_2)$ with $0<f_2<f_1$.
A simple integration leads to
$$
\bar x
  = \left\{\begin{array}{ll}
     1 &~\mbox{for} ~ f < f_2 \bar x \\
     (\rho +a)(1+a)^{-1} &~\mbox{for} ~ f_2 \bar x < f < f_1 \bar x \\
     a(1+a)^{-1} &~\mbox{for} ~ f_1 \bar x < f .
    \end{array} \right.
$$
As $f$ is raised from zero in the system with $f_2 < \rho f_1$,
the first breakdown occurs at $f =f_2$, yielding $\bar
x=(\rho+a)(1+a)^{-1}$, and subsequently the second one at $f
=(\rho +a)(1+a)^{-1}f_1$. For $f_2 > \rho f_1$, on the other hand,
there appears only one breakdown at $f =f_2$. In these two cases
the critical stress $f_c$, beyond which the system breaks down
significantly, is thus given by $(\rho +a)(1+a)^{-1}f_1$ and
$f_2$, respectively.  Note that the survival fraction
$a(1+a)^{-1}$ is negligible for small healing parameter ($a\ll
1$).
\begin{figure}
\centering{\resizebox*{!}{6cm}{\includegraphics{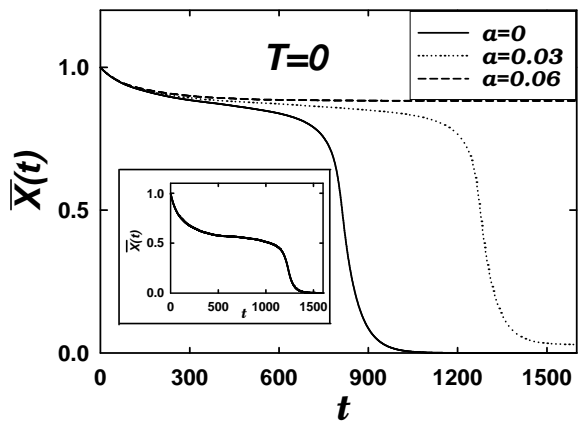}}}
\caption{Evolution of the average fraction $\bar x$ of living
cells with time $t$ (in units of the delay time) for the Gaussian
distribution of $\bar h =1$ and $\sigma =0.2$ at $T=0$. The stress
is $f=0.685$ slightly larger than $f_c (0) \approx 0.677$.  Solid,
dotted, and broken lines correspond to $a=0$, $0.03$, and $0.06$,
respectively. Shown in the inset is the behavior for $\sigma =0.6$
(still at $T=0$ and $a=0$). The stress $f=0.522$ is again slightly
larger than $f_c (0) \, (\approx 0.518$ for $\sigma =0.6$).}
\label{fig:zero}
\end{figure}

We now turn our attention to the time evolution of the system. We
consider first the simplest case of equal load sharing, 
and integrate directly the equation of motion (\ref{xk}) in a numerical way.
We consider the Gaussian distribution of the tolerance with unit mean and standard
deviation $\sigma =0.2$, mostly in a system of $N=10^4$ cells.
Other distributions including the Weibull distribution have also
been considered, only to give essentially the same results.
Specifically, we have used ten different configurations of the
tolerance distribution and set the relaxation time $\tau=50$ and
the time step $\Delta t =0.1$. These parameter values have been
varied, only to give no appreciable difference except for the time
scale.

We first display in fig. \ref{fig:zero} the health status of the
organism at $T=0$, with $f$ being slightly larger than the
critical stress $f_c (a{=}0)\approx 0.677$. As reflected by the
plateau, the system without healing resists the stress for some
time before it breaks down completely. With small amount of
healing, the duration of the plateau becomes longer and there
remains a finite fraction of cells after the breakdown. We refer
to this as the {\em unhealthy} state of the system~\cite{com1}.
Here further increase of $a$ makes the duration of the plateau
extremely long and the system does not break down, implying that
the {\em healthy} state of the organism persists.
For $f$ is smaller than the critical stress $f_c (a{=}0)$, the
average fraction  $\bar x$  of living cells decreases very slowly
(not shown) and the system remains healthy.
Note also that the detailed degeneration behavior of the health status
depends on the tolerance distribution.  For example, a wider distribution in
general brings about severer deterioration initially, followed by long
duration of the unhealthy state before the breakdown (see the inset),
similar to the progression of, e.g., AIDS.

\begin{figure}
\centering{\resizebox*{!}{5.8cm}{\includegraphics{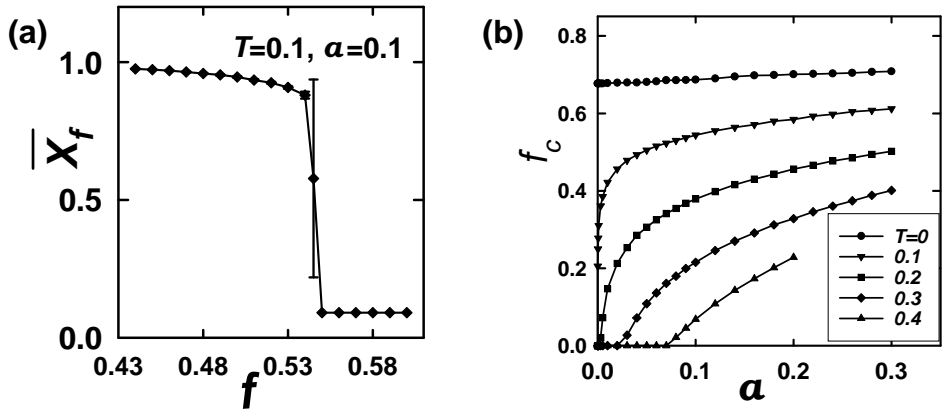}}}
  \caption{(a) Residual fraction $\bar x_f$ of living cells versus $f$ for $a=0.1$ at $T=0.1$. 
(b) Critical stress $f_c$ versus healing parameter $a$ at several
temperatures.}
  \label{fig:fcvsa}
\end{figure}

More interesting results are obtained at finite noise levels.
For $a=0$, non-negligible noise in the system eventually results in
complete breakdown for any finite stress (i.e., $f_c =0$ at
$T\gtrsim 0.1$ although there is numerical ambiguity at very low
noise levels).  When $a\ne 0$, on the other hand, the system may
remain healthy up to some finite stress which depends on $a$,
and thus $f_c$ does not vanish.  To illustrate this, we plot in fig.
\ref{fig:fcvsa}(a) the residual fraction $\bar x$ as a function
of $f$ for $a=0.1$ at $T=0.1$.  It is clearly observed that the system
becomes unhealthy abruptly near $f_c \approx 0.55$ as $f$ is
increased.
Figure \ref{fig:fcvsa}(b) displays how the critical stress
$f_c$ varies with the healing parameter $a$ at several noise
levels. At $T=0$, $f_c$ is observed to increase very slowly with
$a$. In sharp contrast, at low but finite noise levels ($T\ll 1$),
which are presumably the case for biological systems in nature,
$f_c$ increases very rapidly as $a$ is raised from zero, revealing
the crucial role of healing: The presence of even very weak
healing can raise $f_c$ substantially from zero, thus assisting
the system to resist moderate stress. At high noise levels, finite
values of $a$ are necessary to make $f_c$ nonzero.

\begin{figure}
\centering{\resizebox*{!}{8.5cm}{\includegraphics{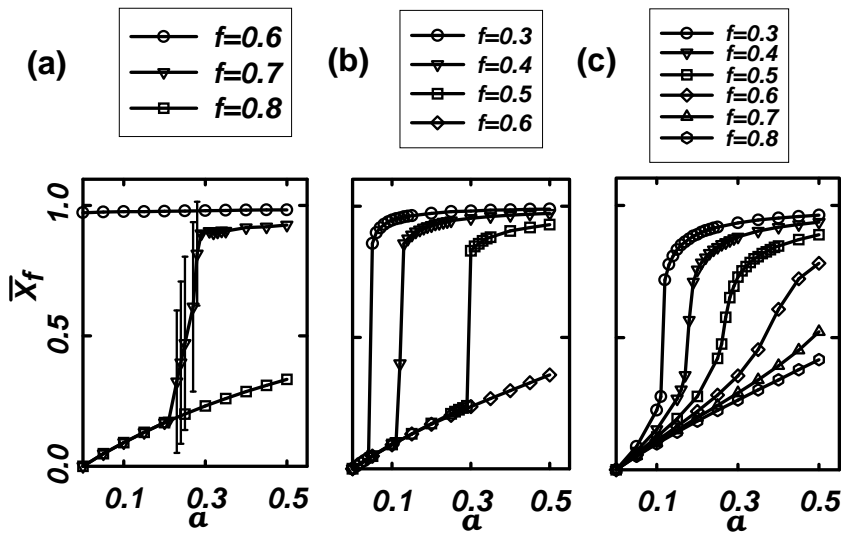}}}
  \caption{Residual fraction $\bar x_f$ of living cells versus healing parameter $a$ for
  several values of the initial stress $f$ at noise level $T=$ (a) 0, (b) 0.2, and (c)
  0.4. Note the large error bars in the ``transition region'' in
 (a). Error bars in other regions are of the order of the symbol size.
  }
  \label{fig:xfvsa}
\end{figure}

%
To further investigate this, we plot in fig. \ref{fig:xfvsa} the
residual fraction $\bar x_f $ of living cells as a function of $a$
for several values of $f$ and $T$.  Figure \ref{fig:xfvsa}(a) shows
$\bar x_f$ as a function of $a$ for three different values of $f$
near $f_c (a=0)=0.677$ at $T=0$.  Observed is that sufficiently strong
stress (i.e., large values of $f$) in general drives the system
out of the healthy state.
Otherwise, as $a$ is raised from zero for given value of $f$,
$\bar x_f$ grows from zero and increases sharply around the
``critical value'' $a_c$ depending on $f$, which is reminiscent of
a phase transition, say, from the unhealthy to healthy state. The
transition appears to become sharper in the presence of noise
($T=0.2$) as shown in fig. \ref{fig:xfvsa}(b), with $a_c$ increasing
with $f$.  The transition, however, tends to be continuous as the
noise level is increased further [see fig. \ref{fig:xfvsa}(c)].

To confirm these findings, we have carried out Monte Carlo simulations
directly and found that the results obtained from 20 independent runs with different
initial configurations display perfect agreement with those from numerical
integrations of the same samples.
In addition, we have also examined the local load sharing case:
Direct Monte Carlo simulations of two-dimensional systems,
with the load being transferred to the shortest paths, show that
qualitative behavior does not change except that the critical stress $f_c$
is smaller than that for the system with equal load sharing.  Further, as the
connectivity of each cell is increased, the behavior approaches that of
the latter system~\cite{com2}.  This provides justification for the use of
equal load sharing in a biological system, which may have enhanced connectivity
in the two- or three-dimensional underlying structure.
%

In summary, we have introduced a dynamic model for failures in
biological organisms and investigated behaviors of the system
under stress and healing.  The dynamics takes into consideration
uncertainty due to imperfections and environmental influences,
described by noise. Such noise has been found to result in
eventual breakdown of the system without healing.  On the other
hand, in the presence of healing, the system in general resists
stress for a longer time compared with the case of no healing, 
and avoids complete breakdown. 
At weak noise the critical stress beyond which
the system breaks down increases rapidly with the healing
parameter, revealing the crucial role of healing. In spite of
simplicity, the model shows quite interesting features, suggestive
of the breakdown phenomena in biological systems.
Finally, we point out that this model is rather general without
details involved and thus provides a convenient starting point for
wide potential applicability. In particular many refinements
toward more realistic descriptions are conceivable.
Corresponding applications to specific biological systems and their implications
are left for further study.

\acknowledgments
This work was supported in part by the 2004 Research Fund of the University of Ulsan.


\begin{thebibliography}{99}

\bibitem{rev}
For a review, see, e.g., \Name{da Silveira R.} \REVIEW{Am. J. Phys.}{67}{1999}{1177};
\Name{Charkrabarti B.K. \and Benguigui L.G.}
\Book{Statistical Physics of Fracture and Breakdown in Disordered Systems}
\Publ{Clarendon, Oxford}
\Year{1997}.
%
\bibitem{hd}
\Name{Daniels H.E.} \REVIEW{Proc. R. Soc. London, Ser
A}{183}{1945}{405}; \Name{Peirce F.T.} \REVIEW{J. Text.
Ind.}{17}{1926}{355}.

\bibitem{asl}
\Name{Anderson J.V., Sornette D., \and Leung K.-t.} \REVIEW{Phys.
Rev. Lett.}{78}{1997}{2140}; \Name{Sornette D.} \REVIEW{J. Phys.
I}{2}{1992}{2089};
\Name{Zapperi S., Ray P., Stanley H.E. \and
Vespignani A.} \REVIEW{Phys. Rev. Lett.}{78}{1997}{1408}.

\bibitem{lt}
\Name{Coleman B.D.} \REVIEW{J. Appl. Phys.}{29}{1958}{968};
\Name{Zhang S.-d.} \REVIEW{Phys. Rev. E}{59}{1999}{1589};
\Name{Newman W.I. \and Phoenix S.L.} \REVIEW{Phys. Rev.
E}{63}{2001}{021507}; \Name{Moral L., Moreno Y., Gomez J.B. \and
Pacheco A.F.} \REVIEW{Phys. Rev. E}{63}{2001}{066106}; \Name{Kun
F., Hidalgo R.C., Hermann H.J. \and Pal K.} \REVIEW{Phys. Rev.
E}{67}{2003}{061802}.

\bibitem{ccy}
\Name{Choi M.Y., Choi J. \and Yoon B.-G.} \REVIEW{Europhys. Lett.}{66}{2004}{62}.

\bibitem{myc}
\Name{Shim G.M., Choi M.Y. \and Kim D.} \REVIEW{Phys. Rev.
A}{43}{1991}{1079}; \Name{Choi M.Y.} \REVIEW{Phys. Rev. Lett.}{61}{1988}{2809};
\Name{Little W.A.} \REVIEW{Math. Biosci.}{19}{1974}{101}.

\bibitem{hp}
\Name{Holford N.H.G. \and Peace K.E.} \REVIEW{Proc. Natl. Acad.
Sci. (USA)} {89}{1992}{11466}.

\bibitem{pc}
\Name{Pradhan S. \and Chakrabarti B.K.} \REVIEW{Phys. Rev.}{E67}{2003}{046124}.
%

\bibitem{com1}
In a real system, however, cells may not be regenerated well in
the unhealthy state, namely, as $\bar{x}$ decreases, so does $a$.
In fact, if $a$ is taken to be this case, the residual fraction
${\bar x}_f$, depending on the specific form of $a$, vanishes or
becomes much smaller than that in the case of constant $a$.

\bibitem{com2}
\Name{Yoon B.-G., Choi M.Y. \and Choi J.} \REVIEW{Unpublished}{}{2005}{}.

\end{thebibliography}
\end{document}